# Interference fading suppression in φ-OTDR using space-division multiplexed probes

Zhiyong Zhao,[1,4] Huan Wu,[2,4] Junhui Hu,[2,3] Kun Zhu,[2,*] Yunli Dang,[1] Yaxi Yan,[2] Ming Tang,[1] and Chao Lu[2]

[1]*Wuhan National Laboratory for Optoelectronics (WNLO), School of Optics and Electronic Information, Huazhong University of Science and Technology, Wuhan 430074, PR China*
[2]*Photonics Research Centre, Department of Electronic and Information Engineering, The Hong Kong Polytechnic University, Hong Kong, PR China*
[3]*Guangxi Key Laboratory of Nuclear Physics and Technology, College of Physics Science and Technology, Guangxi Normal University, Guilin 541004, PR China*
[4]*These authors contributed equally to this work*
[*]*kenny.kun.zhu@gmail.com*

*Last editing date: 10th February 2021.*

**Abstract:** We propose and experimentally demonstrate a novel interference fading suppression method for phase-sensitive optical time domain reflectometry (φ-OTDR) using space-division multiplexed (SDM) pulse probes in few-mode fiber. The SDM probes consist of multiple different modes, and three spatial modes (LP01, LP11a and LP11b) are used in this work for proof of concept. Firstly, the Rayleigh backscattering light of different modes is experimentally characterized, and it turns out that the waveforms of φ-OTDR traces of distinct modes are all different from each other. Thanks to the spatial difference of fading positions of distinct modes, multiple probes from spatially multiplexed modes can be used to suppress the interference fading in φ-OTDR. Then, the performances of the φ-OTDR systems using single probe and multiple probes are evaluated and compared. Specifically, statistical analysis shows that both fading probabilities over fiber length and time are reduced significantly by using multiple SDM probes, which verifies the significant performance improvement on fading suppression. The proposed novel interference fading suppression method does not require complicated frequency or phase modulation, which has the advantages of simplicity, good effectiveness and high reliability.

## 1. Introduction

Phase-sensitive optical time domain reflectometry (φ-OTDR) has attracted a lot of research interests in the last decade, owing to its outstanding performance of distributed vibration sensing, which has shown great potential in various applications, *e.g.* intrusion monitoring, railway transportation monitoring, pipeline safety monitoring and seismic monitoring [1-4], *etc.*

In recent years, to achieve quantitative vibration frequency detection, the optical phase instead of optical amplitude of Rayleigh backscattering signal is usually used in φ-OTDR. It is because the phase change is linear to external vibration therefore ensures high fidelity vibration measurement. The optical phase can be obtained using coherent detection method. However, the coherent detection based φ-OTDR sensors suffer from signal fading issues, including polarization fading and interference fading. Polarization fading results from the polarization mismatch between the Rayleigh backscattered light and the local oscillator. This problem can be solved using polarization diversity detection [5-6]. On the other hand, interference fading originates from coherent Rayleigh interference, because the refractive index distribution of fiber core is not uniform but random, the detected φ-OTDR intensity trace may have some very weak points due to destructive interference. The demodulated phases at very low SNR backscattering points may have severe errors as they are companied with large noise, which eventually leads to frequent false alarms.

Several solutions have been proposed to suppress interference fading. Firstly, multi-frequency pulses method was demonstrated [7], as it was found that the fading positions of φ-OTDR traces are optical frequency dependent, therefore multiple traces that are obtained from different pump light frequencies will be helpful to mitigate misjudgment. Since then, some other frequency-division multiplexed probes methods based on diverse frequency modulation schemes have also been reported, *e.g.* using multiple acousto-optic modulators or an IQ modulator to generate optical pulses with different frequency shifts [8-9]. In addition, inner-pulse frequency-division method has also been proposed, where multiple equivalent probes were obtained from a linear frequency modulated pulse, and the phases were aggregated using rotated-vector-sum method, eventually the fading noises can be removed effectively [10]. This processing method was also adopted by making use of the harmonics of optical intensity modulator [11]. Most recently, spectrum extraction and remix (SERM), as well as spectrum extraction and phase-shift transform methods have also been developed [12-14]. Essentially, all these reported solutions can be categorized as frequency-division multiplexed (FDM) methods. In addition to the FDM probes, another effective interference fading suppression solution is to use phase shift probes, as it turns out that the fading positions of φ-OTDR traces are also different if pulses with varied phase shifts are used [15].

In this work, for the first time, we propose and experimentally demonstrate an interference fading suppression method for φ-OTDR using space-division multiplexed probes. The characteristics of Rayleigh backscattering signal (RBS) of different spatial modes is experimentally investigated using a three-mode fiber. It is found that three modes exhibit unique waveforms with weak correlation. Therefore, different modes can be aggregated for interference fading suppression. The performances of φ-OTDR systems based on single probe and multiple parallel spatial probes are compared qualitatively and quantitatively. A Fading free φ-OTDR system with 4.82 km few-mode sensing fiber and 5 m spatial resolution is demonstrated by aggregating three spatial probes. The proposed solution provides a novel interference fading suppression alternative for φ-OTDR systems, which does not require complicated frequency or phase modulation.

## 2. Experimental setup

To investigate the characteristics of Rayleigh backscattering signals of different spatial modes, a three-mode fiber which supports LP01, LP11a and LP11b modes has been used in the experiments. The refractive index profile of the used three-mode fiber is shown in Fig. 1(a), which has a grade-index distribution. The mode profiles of LP01, LP11a and LP11b are presented in Fig. 1(b).

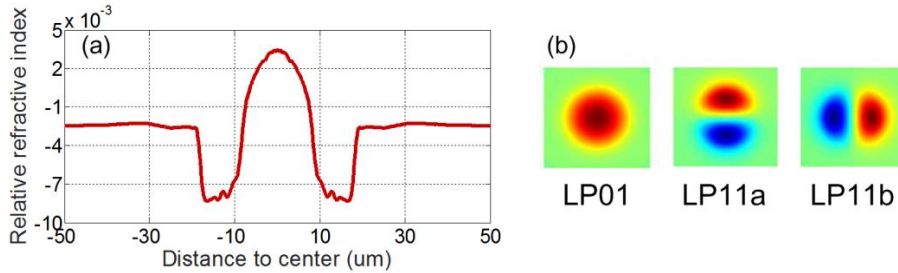

Fig. 1. (a) Refractive index profile of the used three-mode fiber, and (b) are the schematic diagrams of mode profiles of the three modes, respectively.

A heterodyne coherent detection based φ-OTDR system configuration is used in the experiments, whose setup is shown in Fig. 2. A 100Hz narrow linewidth coherent laser (NKT Photonics X15) is used as the optical source, whose output is split into two branches through an optical coupler. The upper branch is used to generate pump pulse through an acousto-optic modulator (AOM, G&H T-M080-0.4C2J-3-F2S) with 80 MHz frequency shift. The used pulse width is 50 ns, which corresponds to 5 m spatial resolution. The optical pulse is then amplified by an erbium-doped fiber amplifier (EDFA, Amonics AEDFA-18-M-FA), which is followed by a 0.8 nm optical bandpass filter to remove the amplified spontaneous emission (ASE) noise. The boosted pulse is then divided into three branches, which will be used for the interrogations of different spatial modes, *i.e.* LP01, LP11a and LP11b modes, respectively. The multiplexing/de-multiplexing of modes is achieved with the help of a photonics lantern [16]. The pulses are launched into the three ports of photonics lantern after passing through three optical circulators, respectively. The used three-mode fiber in the experiment has a length of 4.82 km. On the other hand, the lower branch is served as optical local oscillator, which is also divided into three branches. At the receiver side of each mode, the Rayleigh backscattering light will mix with the local oscillator, and three balanced photodiode (BPD, KY-BPRM-BW-I-FA) are used to detect the beating signals, respectively. The vibration is applied through a speaker with 10 m fiber stick on it at around 4.75 km like the method in [17].The data of all mode channels are collected simultaneously by an oscilloscope (Keysight MSOS404A). The acquired data are processed in digital domain, eventually both the optical intensity and phase of Rayleigh backscattering light can be demodulated.

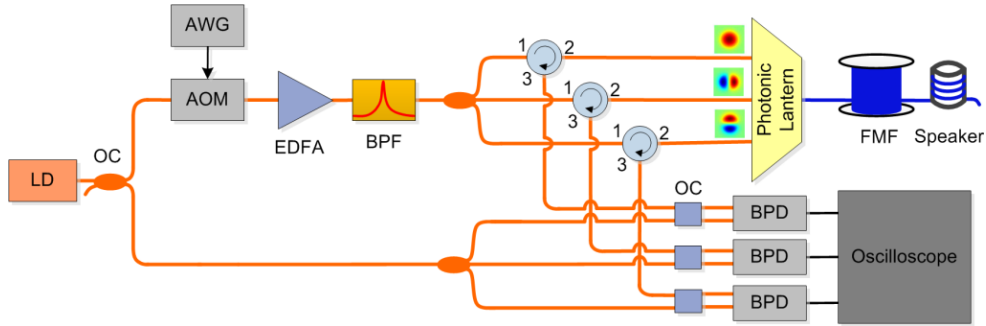

Fig. 2. Experimental setup of the few-mode fiber based φ-OTDR sensing system. LD: laser diode; OC: optical coupler; AOM: acousto-optic modulator; AWG: arbitrary waveform generator; EDFA: erbium-doped fiber amplifier; BPF: band-pass filter; FMF: few-mode fiber; PZT: piezoelectric ceramic transducer; BPD: balanced photo detector.

## 3. Experimental results and discussions

We firstly investigate the characteristics of Rayleigh backscattering signals of different spatial modes. Since the RBS of each mode is detected independently, the RBS intensity and phase information of each mode can be obtained directly using the respective data. Fig. 3(a1), 3(a2) and 3(a3) are the demodulated optical intensity along the whole fiber length of LP01, LP11a and LP11b mode, respectively. It is observed that the optical intensities at many locations are very weak, even lower than the noise floor at some positions due to interference fading. Correspondingly, Fig. 3(b1), 3(b2) and 3(b3) show the demodulated differential optical phase of the three modes respectively, and each contains 100 differential phase traces. Interference fading induced significant phase fluctuations can be observed within the whole fiber length, though only one vibration is applied to the sensing fiber at 4.75 km. Large phase errors will make it impossible to identify the vibrations correctly, and frequent false alarms will make the φ-OTDR system unusable. In order to compare the characteristics of RBS of different spatial modes, the demodulated optical intensities and phases of the three modes are plotted together in Fig. 3(a4) and 3(b4), respectively. Zoom-in view of the traces from 4592 m and

4800 m are presented to show the details of the waveforms. Fig. 3(a4) indicates that the waveforms of RBS traces of diverse spatial modes are different from each other, where the fading points of different modes take place at distinct locations. The correlation coefficients between the modes have been calculated in order to evaluate their correlation. The calculated correlation coefficients are -0.0215 for LP01 and LP11a modes, -0.1473 for LP01 and LP11b modes, 0.1005 for LP11a and LP11b modes, respectively. The differential optical phase traces of different modes presented in Fig. 3(b4) also show the difference between the modes. Due to the severe phase errors caused by interference fading, the reliability of the sensing system is very bad if only one mode is used.

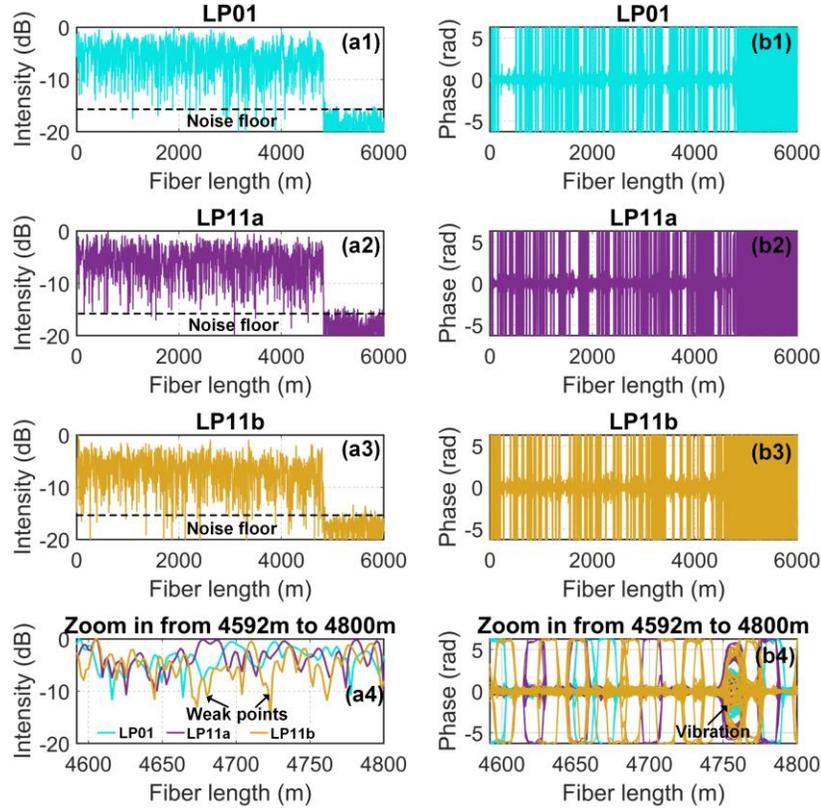

Fig. 3. (a1) - (a3) are the demodulated optical intensities of Rayleigh backscattering signals of the three modes, respectively; (a4) is the zoom-in view of optical intensity traces of the three modes between 4592 m and 4800 m; (b1) - (b3) are the demodulated optical phases of Rayleigh backscattering signals of the three modes, respectively; (b4) is the zoom-in view of optical phase traces of the three modes between 4592 m and 4800 m.

The waveforms in Fig. 3 indicate that the interference fading in φ-OTDR can be addressed by aggregating the RBS of multiple probes that consist of different spatial modes. To verify the feasibility, the Rayleigh backscattering signals of multiple spatial modes based probes are then aggregated using rotated vector sum method to investigate the characteristics of synthesized signals [10]. Fig. 4(a1) and 4(a2) show the demodulated optical intensities of aggregated RBS of two modes based probes, including the combinations of fundamental mode and high-order mode, as well as two high-order modes, *i.e.* LP01+LP11a, LP11a+LP11b, respectively. It is observed that the interference fading induced intensity fluctuation is reduced. The lowest optical intensity of LP01+LP11a aggregated intensity trace is 2.5 dB higher than the noise floor, and it is 1.9 dB for that of the LP11a+LP11b aggregated intensity trace. The less SNR improvement may due to the high correlation between the LP11a and LP11b modes, since they are degenerate modes. Meanwhile, the demodulated differential optical phases of the aggregated RBS of two modes are shown in Fig. 4(b1) and 4 (b2), respectively. Compared with the result of each single probe as shown in Fig. 3, it is seen that the probability of interference fading caused phase errors are reduced notably. In order to further investigate the performance of multiple probes with different spatial modes synthesized φ-OTDR system, all the three modes are used and the demodulated optical intensity after aggregation is shown in Fig. 4(a3). The lowest optical intensity is 5 dB higher than the noise floor, which shows better fading suppression performance in comparison with the case of two probes aggregation. Fig. 4(b3) shows the demodulated differential optical phase after aggregation of three modes simultaneously. It is seen that the interference fading caused phase errors are suppressed completely, which verifies the excellent feasibility of interference fading suppression using multiple space-division multiplexed probes with several modes. To compare the performance of two modes and three modes aggregation, the demodulated intensities after aggregation are plotted in Fig. 4 (a4), and the corresponding differential optical phases are plotted in Fig. 4(b4), where only the zoom-in view of traces between 4592 m and 4800 m are shown. The result indicates that the intensity fluctuation with three modes aggregation is less than that of two modes aggregation. Meanwhile, it also shows that aggregating two modes cannot eliminate fading caused phase errors completely; however, after aggregating three modes, a φ-OTDR system without false alarm can be achieved. The results demonstrate that the fading issue in φ-OTDR can be addressed by aggregating several probes with different modes. Better suppression performance with more modes can be expected.

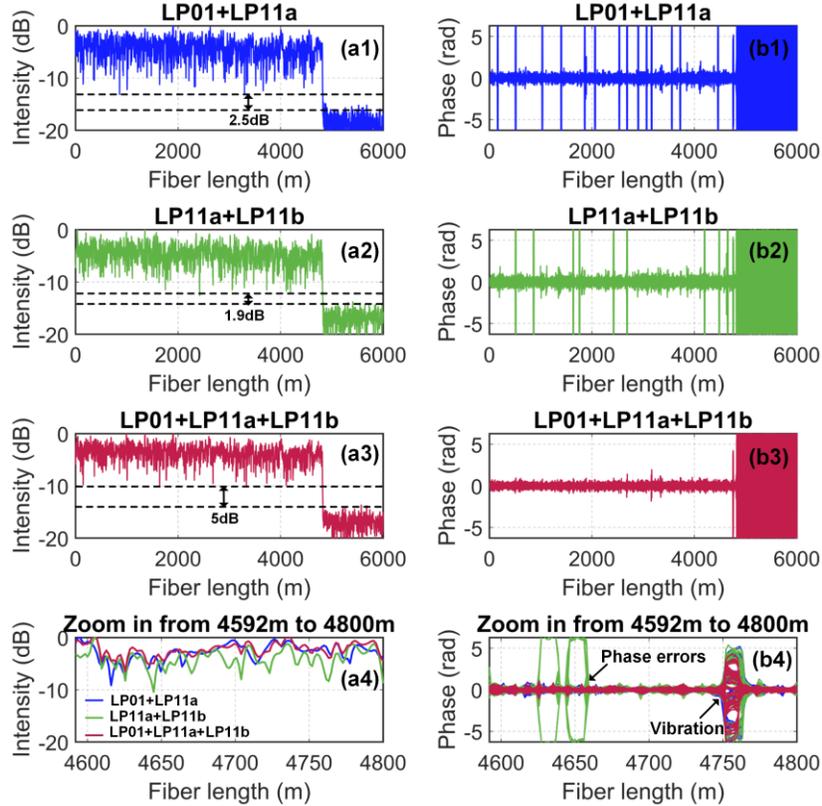

Fig. 4. (a1) - (a3) are the demodulated optical intensities of aggregated RBS of LP01+LP11a, LP11a+LP11b, and LP01+LP11a+LP11b modes, respectively; (a4) is the zoom-in view of optical intensity traces between 4592 m and 4800 m; (b1) - (b3) are the demodulated optical phases of aggregated RBS of LP01+LP11a, LP11a + LP11b, and LP01+LP11a+ LP11b modes, respectively; (b4) is the zoom-in view of optical phase traces between 4592 m and 4800 m.

To quantitatively compare the characteristics of single probe signals and multi-probe aggregated signals, the fading probabilities over fiber length and time are both analyzed statistically. Firstly, histograms of the normalized optical intensity of φ-OTDR traces within one period are calculated, as shown in Fig. 5(a1). For single probe signal (*i.e.* LP01, LP11a or LP11b mode), the histograms exhibit typical Rayleigh distributions with peaks located at around 0.2 (-6.99dB), indicating that many locations tend to occur fading, which matches well with the waveforms in Fig. 3 (a1)-(a3). If two modes are aggregated (*i.e.* LP01+LP11a or LP11a+LP11b), the histograms are obviously right shifted with peaks centred at 0.32 (-4.95dB). This indicates that fading probability will be greatly reduced because the probability of low intensity is decreased. It is worth mentioning that the histograms have no big difference no matter it is the aggregation of one fundamental mode and one high-order mode (*i.e.* LP01+LP11a) or the aggregation of two high-order modes (*i.e* LP11a+LP11b). In addition, the normalized intensity histogram of three modes aggregated signal (*i.e.* LP01+LP11a+LP11b) is also plotted. The centre peak is right shifted to 0.42 (-3.77dB), suggesting that fading probability can be further suppressed by aggregating more probes. Cumulative distribution function (C.D.F.) is then calculated to characterize the six histograms, as shown in Fig. 5(b1). C.D.F. curves are very close if the same number of probe is used, *e.g.* for single probe (LP01, LP11a, LP11b mode) or two probes (LP01+LP11a or LP11a+LP11b). However, with the aggregation of more probes, the C.D.F. curve is more right shifted. Considering using 0.1 (-10dB) as the threshold, the percentages of points below the threshold are 11.72%, 11.18%, 9.72% for LP01, LP11a and LP11b. The percentages are decreased significantly to 0.77% and 0.59% for LP01+LP11a and LP11a+LP11b aggregated signals. For LP01+LP11a+LP11b, there is no point lower than the threshold; therefore a fading free φ-OTDR system is achieved.

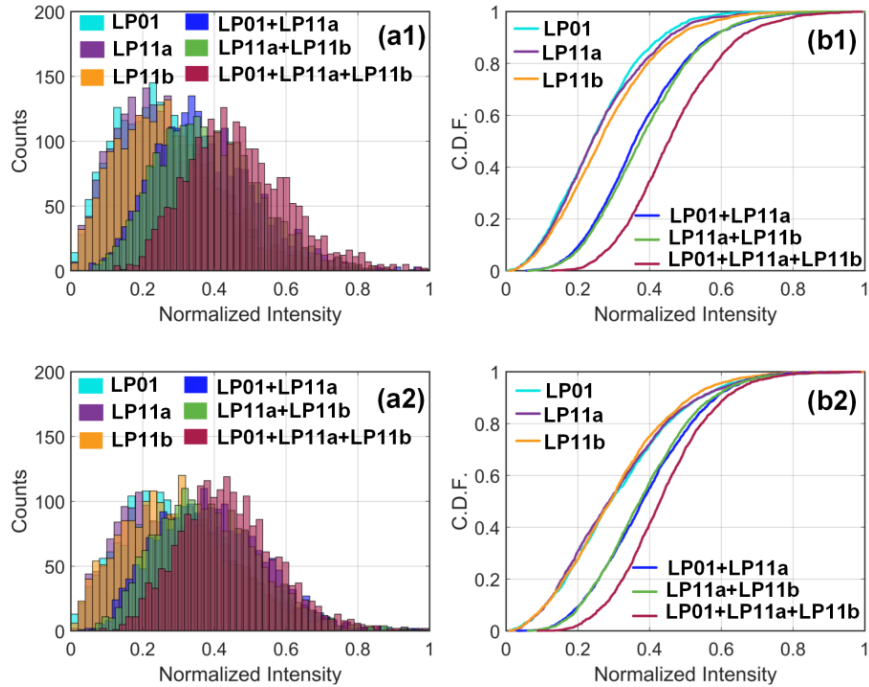

Fig. 5. (a1) is the probability distribution of normalized optical intensity of φ-OTDR trace within one period; (b1) is the cumulative distribution function of normalized optical intensity

of φ-OTDR trace within one period; (a2) is the probability distribution of normalized optical intensity at a random location of 2000 φ-OTDR traces; (b2) is the cumulative distribution function of normalized optical intensity distribution at a random location of 2000 φ-OTDR traces.

Due to the random nature of Rayleigh backscattering, interference fading will not only occur along the fiber, but will also happen along the time. Statistical analysis is also carried out by investigating the fading probability of a random location on the same sensing fiber. We acquired 2000 φ-OTDR traces once every 0.8 s with the total acquisition time of 1600 s. The histograms of normalized intensity over time at 4 km location are drawn in Fig. 5(a2). The statistics results show a similar trend as the statistics over the whole fiber length as presented in Fig. 5(a1). Fig. 5(b2) shows the corresponding C.D.F. of the normalized intensity at 4km location over 2000s. Similar trend can be observed by aggregating more probes; intensities are more likely to be higher values. Again, if 0.1(-10dB) is selected as threshold, the intensity of LP01, LP11a and LP11b mode have respectively 10%, 8.2%, and 8.7% probabilities below the threshold while that of LP01+LP11a and LP11a+LP11b modes both have only 0.65% probability. In addition, LP01+LP11a+LP11b mode exhibit zero probability below the threshold. Thus, both statistical analysis along fiber length and time verify the effectiveness of fading suppression by aggregating multiple modes.

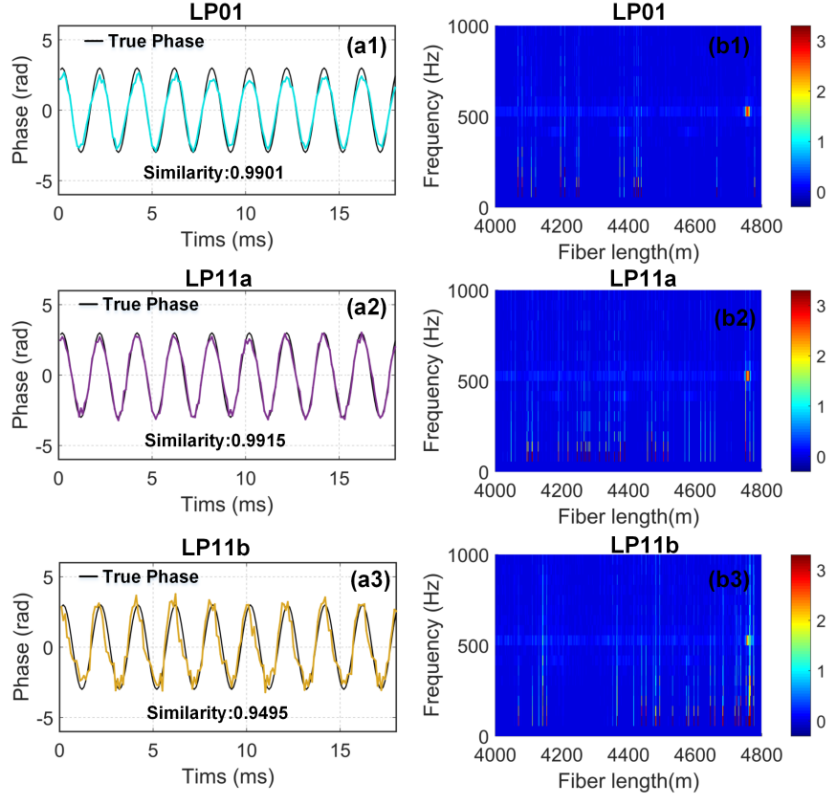

Fig. 6. Vibration detection result using single probe. (a1)-(a3) are the demodulated phases as a function of time at the vibration location; (b1)-(b3) are the extracted vibration frequency spectra along the fiber length.

Apart from fading suppression performance, the vibration detection performance of the sensing system is also investigated. The demodulated phase at the vibration location and the extracted vibration frequency spectrum from single probe of LP01, LP11a and LP11b mode are presented in Fig. 6. No obvious distortion can be observed from the demodulated phases as show in Fig. 6(a1)-(a3), and they all match well with the truly applied vibration signal with similarities better than 0.9495. In addition, it shows that the 500Hz vibration frequency can be extracted correctly at the vibration location. However, significant noises can be observed at some positions in each frequency spectrum (Fig. 6(b1)-6(b3)), which is due to interference fading induced errors. The noises will make it difficult to identify the vibration event correctly but lead to false alarms. It should be mentioned that the weak 500Hz frequency signal along the fiber in these spectra is generated because the mechanical vibration of the speaker is propagated to the 4.8 km main fiber reel.

For comparison, the demodulated phase and the extracted vibration frequency spectra from multiple probes aggregated signal are plotted in Fig. 7. The results indicate that the demodulated phases have good similarity (>0.9816) with the applied vibration signal. Moreover, the interference fading induced noises can be effectively reduced in the frequency spectra, particularly for the three probes aggregated case (Fig. 7(b3)), where the noises are eliminated. The comparison shows significant performance improvement on vibration detection against interference fading induced error using multiple SDM probes simultaneously.

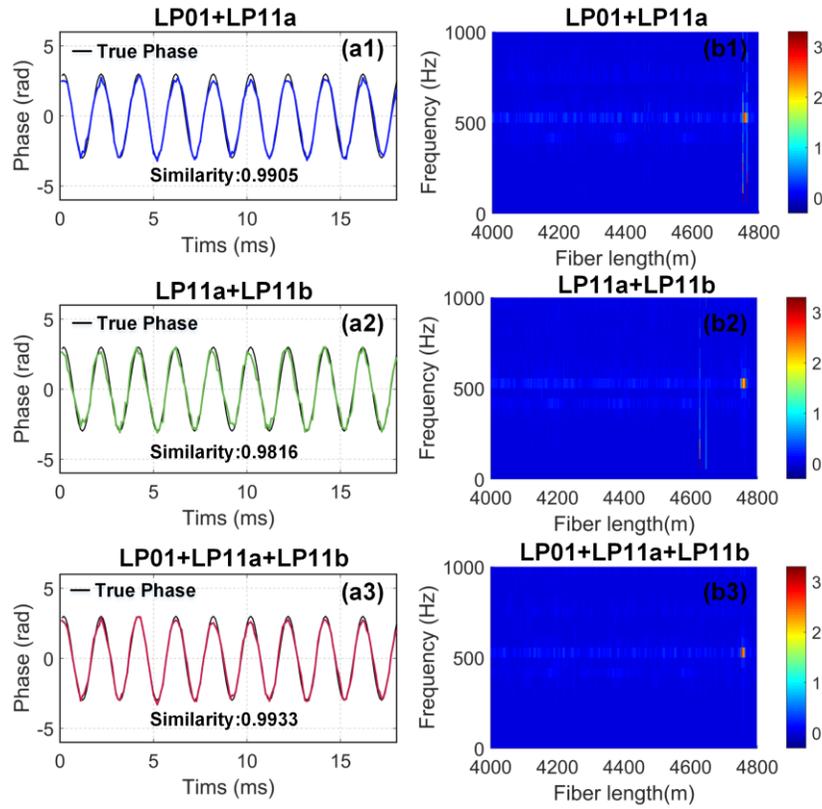

Fig. 7. Vibration detection result using multiple probes simultaneously. (a1)-(a3) are the demodulated phases as a function of time at the vibration location; (b1)-(b3) are the extracted vibration frequency spectra along the fiber length.

## 4. Conclusion

In conclusion, we demonstrate a novel interference fading suppression method of φ-OTDR using space-division multiplexed probes in few-mode fiber, in which the probes consist of distinct modes. It is observed experimentally that the waveforms of φ-OTDR traces of distinct spatial modes are completely different, where the interference fading locations vary among modes. The interference fading suppression effect of multiple probes aggregated signal has also been experimentally verified, and the result indicates that the fading can be suppressed completely using three probes with different modes. In addition, the probabilities of fading over both fiber length and time at a specific position are analyzed statistically, and the results show significant performance improvement on fading suppression using multiple SDM probes. Finally, accurate vibration frequency detection with fading noise eliminated is demonstrated with a 4.85km few-mode sensing fiber and 5m spatial resolution. The proposed SDM probes based sensing system provides a novel simple and effective interference fading suppression method that does not require complicated frequency or phase modulation.

**Funding**

National key R&D Program of China (2018YFB1801002); The Hong Kong Research Grants Council (PolyU 152168/17E); The Hong Kong Polytechnic University (YW3G, ZVGB); Guangxi key R&D program (AB18221033); Guangxi one thousand Young and Middle-aged College and University Backbone Teachers Cultivation Program (20181227).

**Disclosures**

The authors declare no conflicts of interest.